\documentclass{article}
\usepackage{waspaa19,amsmath,graphicx,url,times}
\usepackage{caption, verbatim, cite, multirow, subcaption}
\usepackage{color}
\usepackage{commath,verbatim}

\title{Room Geometry Estimation from Room Impulse Responses\\
 using Convolutional Neural Networks}

\name{Wangyang Yu,$^{\star}$ \qquad W. Bastiaan Kleijn$^{\dagger\star}$}

\address{$^{\star}$ EEMCS, Delft University of Technology, Mekelweg 4, 2628 CD, Delft, The Netherlands \\
    $^{\dagger}$ ECS, Victoria University of Wellington, Kelburn, Wellington 6012, New Zealand} 

\begin{document}

\ninept
\maketitle

\begin{sloppy}

\begin{abstract}
We describe a new method to estimate the geometry of a room given room impulse responses. The method utilises convolutional neural networks to estimate the room geometry and uses the mean square error as the loss function. In contrast to existing methods, we do not require the relative positions of sources and receivers in the room. The method can be used  with only a single room impulse response between one source and one receiver. The proposed estimation method can achieve an average of six centimetre accuracy. In addition, the proposed method is shown to be computationally efficient compared to state-of-the-art methods.
\end{abstract}


\section{Introduction}
\label{sec:intro}

Augmented reality (AR) is an immersive audio-visual environment where artificial objects are added to a real-world scenario, providing the user with an enhanced and interactive experience \cite{arwiki}. Augmented reality will play an increasingly important role in numerous contexts, such as education, manufacturing, and archaeology. An accurate description of the acoustic environments is essential for generating perceptually acceptable sound in an AR system. Room geometry is an important aspect of  modelling an acoustic environment accurately. In this paper we consider the estimation of room geometry from room impulse responses as that facilitates a quick and practical measurement.

The room impulse response, the transfer function between the sound source and the listener, characterises the acoustics environment of a room. It is composed of direct-direction sound, early reflections, and late reverberation. A room impulse response is affected by the position of the sound source and the receiver, the room geometry and the reverberation time. The image-source method \cite{kuttruff2014room, allen1979image} is commonly used to model reflections in a room.

Existing algorithms to estimate room geometry from room impulse responses all require prior information about the configuration of the sources and the microphones \cite{Dokmani12186,7471691,7471625, 6811439}. \cite{6811439} uses room impulse responses and a set of time of arrival (TOA) measurements to estimate 2D room geometry. It assumes that the TOA measurements are labelled and room impulse responses consist of direct sound and the first and the second order reflections. \cite{Dokmani12186} proposes a method to estimate the 3D room shape from room impulse responses by exploiting the properties of Euclidean distance matrices and the first order reflections. Although it requires only a single source, it requires at least four receivers and their pairwise distances. In addition, it may misclassify higher order reflections as the first order reflections \cite{7471691}. In \cite{7471691}, the room geometry is estimated from one sound source by a two-step intuitive geometrical method. The proposed method requires five receivers and their pairwise distances. \cite{7471625} infers the room geometry efficiently by labelling echoes and inferred image source positions by labelled echoes, which requires at least two sources and five receivers. 

In contrast to the above mentioned state-of-the-art methods, we would like to estimate the room geometry and without knowledge of positions of source and receiver. Averaging the outcomes for different responses can then be used to increase estimation accuracy. Since a room impulse response contains information about the room geometry and the geometry information is independent of positions of sources and receivers, we hypothesise estimating room geometry does not require multiple sources or receivers or their positions. A natural solution to the problem of estimating room geometry from a single response is a method based on machine learning. 

In convolutional neural networks (CNNs), the receptive field of each neuron is processed with a kernel that does not vary across the input data. For our geometry-estimation problem, this corresponds to assuming that the room impulse response has a similar structure across all delays. CNNs were first proposed by \cite{FUKUSHIMA1988119} for visual pattern recognition. As a result of the increased computational power and the availability of large databases, CNNs have seen a rapid increase in usage in recent years. Many variations of CNN architectures have been developed, such as AlexNet \cite{Krizhevsky:2012:ICD:2999134.2999257} and VGG-16 \cite{DBLP:journals/corr/SimonyanZ14a}. In addition, CNNs have been used for various applications such as image classification \cite{krizhevsky2012imagenet,7305792,7973178}, speech recognition \cite{6857341,8384470,7552554}. Recent applications, such as reverberation time estimation \cite{7974586}, confirm that CNNs are able to show a good modelling ability for acoustic problems and outperform state-of-the-art algorithms in this context, which motivates us to use CNNs for our problem. 

The main contribution of our paper is the usage of convolutional neural networks to estimate room geometry. In contrast to state-of-the-art methods for room geometry estimation, our method does not require the position or distance of receivers and sources. In its basic form, the method requires only one room impulse response between a single sound source and a single receiver.  The proposed method is computationally efficient compared to state-of-art algorithms.

This paper is organised as follows. In section 2, we formulate our room geometry estimation problem and describe the network architecture. In section 3, we describe how we generate the database, discuss our experimental setup and analyse the results. Finally, we conclude our paper in section 4.

\section{CNN based Room Geometry Estimation}
\label{sec:cnn}

We use CNNs to estimate room geometry from room impulse responses. The room impulse response depends on reflection coefficients and room geometry. We define room geometry to be a three-dimensional vector, which contains the length, width, and height of a room. We consider reflection coefficients as a nuisance factor in our problem.  In this section, we first describe the architecture of and metrics for our CNN model. Then we propose a method to improve accuracy. Finally, we discuss the effect of reflection coefficients and reverberation time.

\subsection{Architecture and metrics}

Convolutional neural networks are considered a powerful modelling technique in various applications \cite{krizhevsky2012imagenet,7305792,7973178,6857341,8384470,7552554}. Furthermore, a CNN generalises a filter with the activation function. CNNs contain a set of generalised filters of different levels to extract the various features from the signals. By each convolutional operation, each signal sample is processed with a filter which does not vary across the signal. The parameters of each filter are learned through the training process. CNNs can thus learn features of a signal. Room geometry is an underlying feature of room impulse responses and we assume room impulse responses show a certain structure with room geometry across all delays. Consequently, applying CNNs on room impulse responses is expected to extract room geometry information.

Since the room geometry is described by continuous variables, we formulate the room geometry estimation problem as a regression problem, where the model can output the continuous room geometry estimates directly. To solve the problem, our neural network has three output nodes for the length, width and height of a room. Our model takes one room impulse response in the time domain as the input, without any pre-processing. The network estimates the geometry of a room for each room impulse response of the given room. Since the ordering of the three lengths of the geometry is arbitrary, we must re-order the geometry vector in ascending order in a pre-processing step.

We adopt a commonly used CNN architecture as a basis. In this architecture each convolutional layer is followed by a batch normalisation layer \cite{DBLP:journals/corr/IoffeS15} and an activation function. Since our input signal is the raw time-domain signal, we use one-dimensional convolutional layers and one-dimensional batch normalisation layers. To keep a balance between the number of parameters and the modelling ability of neural networks, the neural network consists of six one-dimensional convolutional layers and two fully connected layers. The number of the filters in the convolutional layers increases with depth because the output dimensionality of the convolutional layers decreases. We use a rectified linear unit (ReLU) activation function \cite{Nair:2010:RLU:3104322.3104425} as the activation function. To prevent overfitting, early stopping is used for regularisation in our neural network \cite{earlystop}. Early stopping is performed when the validation performance degrades in $S$ successive epochs. Our network architecture and corresponding parameters are shown in Table \ref{tab:cnn}, where $n$ denotes the batch size. The network contains $178413$ trainable parameters in total.  
\begin{table}[h]
\scriptsize
\centering
\caption{Network Architecture}
\label{tab:cnn}
\begin{tabular}{| c | c | c | c | c |}
\hline
Operation & Kernel Size & Stride & $\#$ filters & Output Size \\
\hline
Input & & & & $(n,4096)$ \\
\hline
Reshape & & & & $(n, 1, 4096)$\\
\hline
Conv1D & $4$ & $4$ & $10$ & $(n, 10, 1024)$ \\
\hline
BatchNorm1D & & & & $(n, 10, 1024)$\\
\hline
ReLU & & & & $(n, 10, 1024)$\\
\hline
Conv1D & $4$ & $4$ & $20$ & $(n, 20, 256)$ \\
\hline
BatchNorm1D & & & & $(n, 20, 256)$\\
\hline
ReLU & & & & $(n, 20, 256)$\\
\hline
Conv1D & $4$ & $4$ & $40$ & $(n, 40, 64)$ \\
\hline
BatchNorm1D& & & & $(n, 40, 64)$\\
\hline
ReLU & & & & $(n, 40, 64)$\\
\hline
Conv1D & $4$ & $4$ & $80$ & $(n, 80, 16)$ \\
\hline
BatchNorm1D & & & & $(n, 80, 16)$\\
\hline
ReLU & & & & $(n, 80, 16)$\\
\hline
Conv1D & $4$ & $4$ & $160$ & $(n, 160, 4)$ \\
\hline
BatchNorm1D & & & & $(n, 160, 4)$\\
\hline
ReLU & & & & $(n, 160, 4)$\\
\hline
Conv1D & $4$ & $4$ & $160$ & $(n, 160, 1)$ \\
\hline
BatchNorm1D & & & & $(n, 160, 1)$\\
\hline
ReLU & & & & $(n, 160, 1)$\\
\hline
Reshape & & & & $(n, 160)$\\
\hline
Fully connected & & & & $(n, 40)$\\
\hline
Fully connected & & & & $(n, 3)$\\
\hline
\end{tabular}
\end{table}

The mean square error is used as our loss function to minimise the squared distance between the estimated room geometry and the true room geometry. The loss function for each batch is defined as
\begin{equation}
\text{MSE}(L, \hat{L}) = \frac{1}{n}\displaystyle\sum_{i=1}^n(L(i,:)-\hat{L}(i,:))^{\circ 2},
\end{equation}
where $n$ denotes the batch size, $L$ is a $n$ by $3$ matrix which denotes the true room geometry per batch, $\hat{L}$ denotes the corresponding estimated room geometry matrix, and $(\cdot)^{\circ 2}$ denotes Hadamard power. We chose the mean square error loss since it is relatively sensitive to outliers, which we would like to suppress in the room geometry estimation problem. 

The network is trained with the Adam optimiser, a robust stochastic gradient-based optimisation algorithm \cite{DBLP:journals/corr/KingmaB14}, to minimise the training loss. Compared to other optimisation algorithms, the Adam optimiser generally converges faster for problems with a large amount of data and parameters, making it well-suited to our estimation problem. 

To characterize the estimation performance of our method, we evaluate median, bias and variance on the test data. Median measures the central tendency of estimation error and it is less skewed by imbalanced distribution than the mean. In statistics, bias measures the mean deviation of our estimates from the true value and variance measures how much our estimates vary from the mean estimated value. Minimising mean square error is able to keep balance between bias and variance since the relationship between mean square error, bias and variance can be described as
\begin{equation}
\text{MSE} = \text{Bias}^2 + \text{Variance}.
\end{equation}
Since bias is also a parameter that a CNN model tries to learn during the training process, a CNN model should in principle result in an unbiased estimator.  For an unbiased estimator, we can increase the precision by averaging over the estimates. 

\subsection{An improved algorithm}

Assuming the estimates are not strongly biased, we propose an improved algorithm for room geometry estimation. For each room, we select $N$ random independent room impulse responses. The method is to average over the $N$ estimates to calculate the final estimate for the room. The variance of the estimator will decrease by averaging over $N$ independent estimates. Although the accuracy is limited by the bias, the estimation precision can be increased.

\subsection{The effect of reflection coefficients and reverberation time}

In addition to the room geometry, the room impulse responses are also affected by reflection coefficients. We aim to investigate if fixed or varying reflection coefficients have an effect on the accuracy of room-geometry estimation. We hypothesise that fixed reflection coefficients result in a more accurate estimate.

Sabine's formula commonly quantifies reverberation time:
\begin{equation}\label{sab}
RT_{60} = \frac{24 ln 10}{c_{20}} \frac{V}{Sa} \approx 0.1611 \text{sm}^{-1}
\frac{V}{Sa},
\end{equation}
where $c_{20}$ is the speed of the sound in the room for $20$ degrees Celsius, $V$ is the room volume, $S$ is the total surface area of the room and $a$ is the average absorption coefficient of room surfaces. From \eqref{sab}, we can conclude that reverberation time is related to room geometry and reflection coefficients. We expect that varying reverberation time affects the accuracy of estimation. 

\section{Experiments}
\label{sec:experiment}

In this section, we present our experiments. In the first subsection, we describe how we generated our database for training and testing. We describe the setup our experiments in the second subsection. Finally, we show and analyse our experimental results.

\subsection{Database Generation}

We need a large-scale dataset of good quality to train our deep neural networks. To build such a dataset, we used the image-source method to simulate RIRs \cite{habets2014}. We assume the room is shoe box shaped. The speed of sound was set to $c = 340$ m/s. The sampling frequency was set to $8000$ Hz. The length of each RIR was $4096$, corresponding to approximate $0.5$ seconds. It contains at least the direct path signal and early reflections in an indoor environment.  Each dimension of room geometry, i.e., length $\times$ width $\times$ height, was assumed to be independently uniformly distributed between $6 \times 5 \times 4$ m and $10 \times 8 \times 6$ m.  In each room, we randomly generated $16$ RIRs since it is a multi-to-one mapping and we found it outperforms other cases in our preliminary tests. However, each room impulse response will be processed independently. The validation dataset and the test dataset are generated in the same way as the training dataset. In our experiment, the size of the training dataset was $336000$ RIRs, the size of the validation dataset was $96000$ RIRs, and the size of the test dataset was $48000$ RIRs. 

\subsection{Experimental Setup}

In this subsection, we describe how we set up our experiments. We first discuss the experiments for reverberation time and error analysis. Then we discuss the setup of the experiments for improved methods. Finally, we describe the general experimental setup.

Our first experiment was to determine the effect of reflection coefficients and reverberation time. We divided our experiments into two cases, fixed and varying reflection coefficients. We first fixed the reflection coefficients and generated the database on this randomly generated set of reflection coefficients. In this setup, the varying reverberation time is only related to the change of room geometry. We then removed this restriction. The varying reflection coefficients database were generated to guarantee the reverberation time uniformly distributed between $0.4$ s and $1$ s, a range chosen to be representative of real-world environments. After that, we compared the performance of these two cases in terms of mean square error, bias and variance in test set. We use the mean estimation error to approximate bias. Our further experiments were based on varying reflection coefficients.

To facilitate interpretation of our results we include an error analysis. To begin with, we plotted the error distribution in the test set, where we mix the mean square error for length, width and height because we want to see an overall performance on these three elements. For further analysis, we generated eight rooms, each with a random reverberation time. In each room, we randomly placed $100$ sources and $100$ receivers and calculated the room impulse responses among them. We also plotted the mean square error distribution in the eight rooms. In addition, we plotted estimation error performance in each room and analysed the result. We compared the error performance in each room to determine if the bias is constant in different room configurations and evaluate how much the estimates deviate from the ground truth in each room.

We proposed an improved method to increase the estimation accuracy. We aim to use experiments to investigate the accuracy we can reach and the effect of the number of estimates. We use the test dataset to do the experiments. We first shuffled and loaded the dataset to compute the estimates of each RIR. Next we reordered the estimates by the true room geometry and group the estimates to one, four, eight and 16 estimates per room to perform average method. Finally, we computed the MSE, median, bias and variance of the average method.  The median is computed with the absolute estimation error.

We used PyTorch to implement our neural network and perform training. We used the default initialisation method in PyTorch. We used a GPU node to train our neural network. The batch size was set to be $50$. The learning rate of Adam optimiser was $0.001$ and the coefficients used for computing running averages of the gradient and its square were set to be $(0.9, 0.999)$. We iterated for $2000$ epochs and recorded the mean square error loss for each epoch. In each epoch, we set the model on evaluation mode and compute the validation error for early stopping monitor. We set $S$ to $30$ which guarantees the training performance without overfitting.  After training, we set the model on the evaluation mode and computed the test error. In addition, we recorded the running time for the geometry estimation of each room.

\subsection{Experimental Results}

In this subsection, we show and analyse our experimental results. We first compare the results of fixed and varying reflection coefficients. Then we show the results of error analysis and plot the error distribution. After that, we discuss the proposed improved method. Finally, we compare our estimation error and running time with one of the traditional methods.

To begin with, we show the mean squared error, bias and variance on fixed and varying reflection coefficients in Table \ref{tab:exp1}. The positive sign indicates our prediction is larger than the true geometry value. The mean squared error, bias and variance show different values with respect to length, width and height because the range on these three elements is different. We find the error is smaller when the reflection coefficients are fixed. This proves that the varying reflection coefficients is a nuisance factor in our estimation problem and varying reflection coefficients has an effect on the accuracy of estimation. From another point of view, the small bias vector confirms that our CNN model is not significantly biased after training and the small variance confirms that most estimation errors are relatively small and they do not vary much. Our remaining experimental results refer to the varying reverberation time condition since it is representative of real-world environments. 

\begin{table}[h]
\scriptsize
\centering
\caption{Mean square errors, bias and variance of fixed and varying reflection coefficients}
\label{tab:exp1}
\begin{tabular}{| c | c | c |}
\hline
Reflection coefficients& Fixed  & Varying\\ 
\hline
MSE (m$^2$) & $[0.005, 0.017, 0.006]$ & $[0.015, 0.018, 0.007]$ \\ \hline
Bias (m) & $ [0.011, 0.001, -0.002] $ & $[0.019, -0.002, 0.004]$ \\ \hline
Variance (m$^2$) & $[0.005, 0.017, 0.006]$ & $[0.015, 0.018, 0.007]$ \\ \hline
\end{tabular}
\end{table}

Next we show the results of the error analysis. The error distribution in the test set and eight selected rooms is shown in Figure \ref{fig:exp2}. Observing the error distribution in the test set, the error follows a long-tailed distribution, which confirms that most estimation errors are relatively small, which is consistent with the small variance in the test set. The error distribution in the eight random rooms approximately follow the same distribution as the test set but the proportion of small errors is different for different rooms. In Figure \ref{fig:exp3}, we plotted the mean estimation error of length, width and height in the eight generated rooms and the error bar represents the corresponding standard deviation. Comparing the mean estimation error of each room, we found there will be a bias for each room and the sign varies with the room. In addition, the standard deviation is also different in different rooms. This indicates that estimation errors mainly result from specific room configurations. Consequently, some room configurations may outperform others.

\begin{figure}
\begin{subfigure}[t]{.5\linewidth}
    \centering\includegraphics[width=\linewidth]{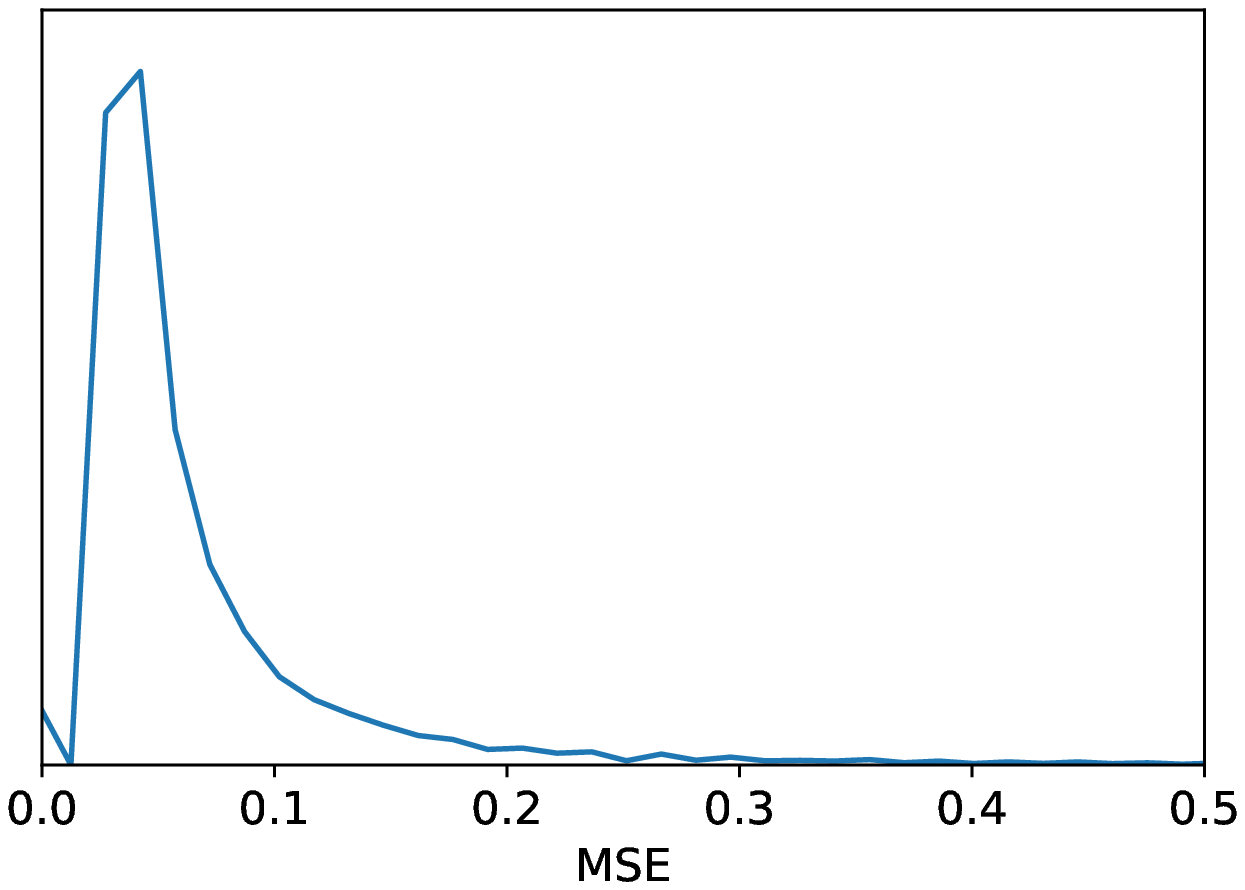}
    \caption{Test set.}
  \end{subfigure}
  \begin{subfigure}[t]{.5\linewidth}
    \centering\includegraphics[width=\linewidth]{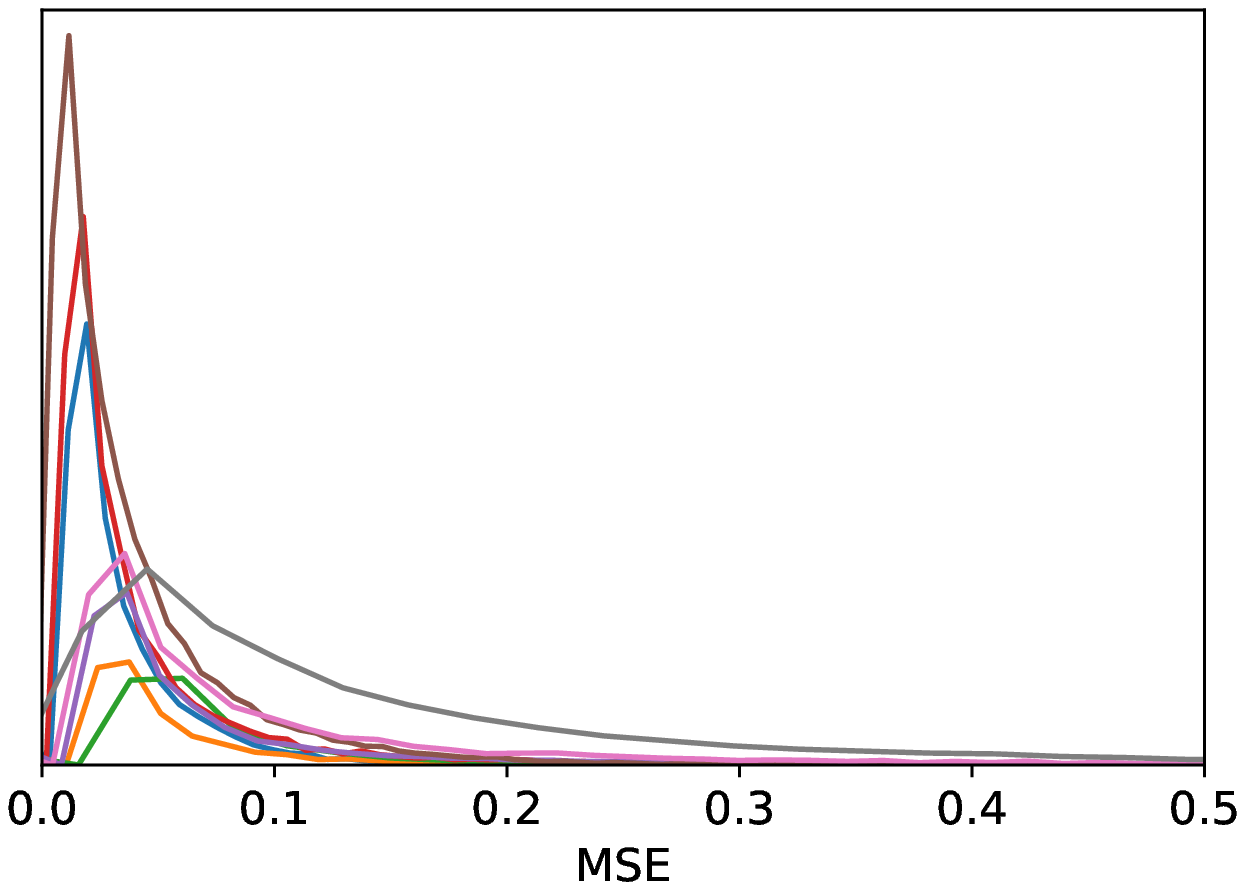}
    \caption{Eight different rooms.}
  \end{subfigure} 
\caption{MSE distribution under varying reflection coefficients.}
\label{fig:exp2}
\end{figure}

\begin{figure}[h]
\centering
\includegraphics[width=0.35\textwidth]{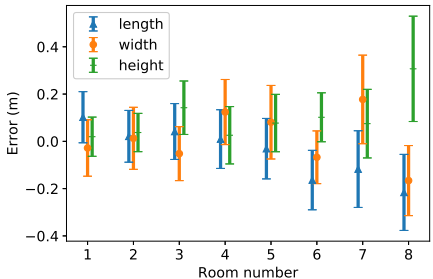}
\caption{Mean estimation error and the standard deviation in eight rooms under varying reflection coefficients.} 
\label{fig:exp3}
\end{figure}

After analysing the test error, we compared the improved estimation method with our baseline method. The bias of estimation error is $[0.019, -0.002, 0.004]$, which does not change by averaging over $N$ estimates. The mean squared error, median and variance under different number of room impulse responses are listed in Table \ref{tab:exp7}. The method with one room impulse response corresponds to our baseline method. From Table \ref{tab:exp7}, we can conclude that the averaging method outperforms our baseline method. The variance decreases via averaging but not decreases by a factor of $N$ since there exists a nuisance factor, reflection coefficients, which makes the room impulse responses in each room are not independent conditioned on room geometry. Since our error distribution is long-tailed, median can represent the central tendency and is less affected by the small portion of large errors. Observing Table \ref{tab:exp7}, median decreases via averaging method. To conclude, the performance is better when more room impulse responses are used for averaging although our estimation is still biased.
\begin{table}
\scriptsize
\centering
\caption{Mean squared error, bias and variance of improved method under varying reflection coefficients.}
\label{tab:exp7}
\begin{tabular}{| c | c | c | c | c |}
\hline
$\#$ RIRs & MSE (m$^2$) & Median (m) & Variance (m$^2$)\\ \hline
$1$ & $[0.015, 0.018, 0.007]$ &  $[0.072, 0.069, 0.044]$ & $[0.015, 0.018, 0.007]$   \\ \hline
$4$ & $[0.006, 0.007, 0.005]$ &  $[0.046, 0.045, 0.027]$ & $[0.005, 0.007, 0.005]$   \\ \hline
$8$ & $[0.004, 0.005, 0.004]$ &  $[0.039, 0.038, 0.022]$ & $[0.004, 0.005, 0.004]$   \\ \hline
$16$ & $[0.003, 0.004, 0.004]$ &  $[0.035, 0.032, 0.017]$ & $[0.003, 0.004, 0.004]$   \\ \hline
\end{tabular}
\end{table}


Finally, we compared our improved method with the method proposed in \cite{7471625} in terms of system requirements, estimation error and average run time. For calculating the run time, the experiments were run on a MacBook Pro Mid 2014 with 2.6 GHz Intel Core i5 processor in Python 3.6.5 and averaged over $3000$ experiments. We average over the square root of the mean square error as our average error for comparison. The result is shown in Table \ref{tab:exp4}. The method in \cite{7471625} uses five sources and five receivers under $96$ kHz sampling frequency while the proposed method only requires sixteen random RIRs under $8$ kHz sampling frequency. From the experimental results, on the one hand, the traditional method performs approximately three times better in terms of average estimation error. On the other hand, in terms of average run time, our CNN based method performs $10^{4}$ better than the method proposed in \cite{7471625}. To conclude, our CNN based room geometry estimation method is computationally efficient with acceptable estimation error and does not require prior knowledge or knowledge of the measurement configuration compared to traditional methods.
\begin{table}
\centering
\caption{Comparison of proposed method and state-of-art method.}
\label{tab:exp4}
\begin{tabular}{| c | c | c |}
\hline
  & Proposed method & Method in \cite{7471625} \\ \hline
Average error (m) & $0.0611$  & $0.0235$ \\ \hline
Average run time (s) & $3.22 \times 10^{-4}$ & $2.43$ \\ \hline
\end{tabular}
\end{table}
 
\section{Conclusions}
\label{sec:conclusion}

In this paper, we used convolutional neural networks to estimate room geometry. We formulated our problem as a regression problem with the mean square error as a loss function. The proposed method only requires one random room impulse response between a single source and a single receiver and does not require knowledge of positions or relative distance. Moreover, we proposed an improved method to increase estimation accuracy. With our proposed method, we can arrive at an average of $[0.064, 0.059, 0.060]$ m estimation accuracy and the median of the absolute estimation error is $[0.035, 0.032, 0.017]$ m. In addition, our method is computationally efficient. A natural extension of our work will focus on more advanced error analysis to determine the reason for the bias and approaches that minimize bias.

\bibliographystyle{IEEEtran}
\bibliography{refs}

\end{sloppy}
\end{document}